\documentclass[aps,prd,twocolumn,nofootinbib]{revtex4-1}
\usepackage{epsfig}
\usepackage{bm}
\usepackage{array}
\usepackage{color}
\usepackage{amsmath}
\usepackage{graphicx}
\usepackage[colorlinks,linkcolor=blue,anchorcolor=blue,citecolor=blue,urlcolor=blue,breaklinks=true]{hyperref}
\usepackage[usenames]{xcolor}
\usepackage[normalem]{ulem}

\begin{document}

\author{Q.~Meng$^{1}$}
\author{E.~Hiyama$^{2,3,4}$}
\author{A.~Hosaka$^{5,4}$}
\author{M.~Oka$^{4,3}$}
\author{P.~Gubler$^{4}$}
\author{K.U.~Can$^{6,3}$}
\author{T.T.~Takahashi$^{7}$}
\author{H.S.~Zong$^{1,8,9,10}$}

\address{$^1$Department of Physics, Nanjing University, Nanjing 210093, China}
\address{$^2$Department of Physics, Kyushu University, Fukuoka 819-0395, Japan}
\address{$^3$Nishina Center for Accelerator-Based Science, RIKEN, Wako 351-0198, Japan}
\address{$^4$Advanced Science Research Center, Japan Atomic Energy Agency, Tokai 319-1195, Japan}
\address{$^5$Research Center for Nuclear Physics (RCNP), Osaka University, Ibaraki 567-0047, Japan}
\address{$^6$CSSM, Department of Physics, The University of Adelaide, Adelaide SA 5005, Australia}
\address{$^7$National Institute of Technology, Gunma College, Gunma 371-8530, Japan}
\address{$^8$Department of Physics, Anhui Normal University, Wuhu, 241000, China}
\address{$^9$Nanjing Proton Source Research and Design Center, Nanjing 210093, China}
\address{$^{10}$Joint Center for Particle, Nuclear Physics and Cosmology, Nanjing, 210093, China}

\title{Stable double-heavy tetraquarks: spectrum and structure}

\begin{abstract}
Bound states of double-heavy tetraquarks are studied 
in a constituent quark model. 
Two bound states are found for isospin and spin-parity $I(J^P) = 0(1^+)$ in the $bb\bar u \bar d$ channel.
One is deeply bound and compact made of colored diquarks, while the other is shallow and extended as a $BB^*$ molecule. 
The former agrees well with lattice QCD results.
A systematic decrease in the binding energy is seen by replacing
one of the heavy quarks to a lighter one.  
Altogether we find ten bound states.  
It is shown for the first time that hadrons with totally different natures emerge from a single Hamiltonian.   
\end{abstract}

\maketitle

This paper addresses the spectrum and structure of bound tetra-quark states with two heavy quarks.
This is a key problem of quantum chromodynamics (QCD) that will lead to resolving confusion in interpreting the nature of exotic resonances observed recently at high energy accelerator facilities 
such as LHC, KEK and BEP.

Hadron spectroscopy has turned to a new phase in the past 15 years
by successive discoveries of new hadron resonances, such as $X(3872)$, $P_c$ 
and others~\cite{Choi:2003ue,Aaij:2019vzc,Hosaka:2016pey}. 
As they do not fit into the conventional meson ($q \bar q$) and baryon ($qqq$) pictures,
their structure and dynamics must be different from the ordinary hadrons.
Another notable feature of some of the newly found resonances is their narrow widths,
in spite of sizable phase space open to hadronic decay channels. 

Various interpretations have been given for the observed exotic 
resonances~\cite{Chen:2016qju,Yamaguchi:2019vea}.  
Some are consistent with loosely bound states of hadrons, forming hadronic molecules.
For instance, $X(3872)$ was suggested to be a molecular bound state of 
$D$ and $\bar D^*$~\cite{Tornqvist:1993ng}.
Another interpretation 
is a threshold cusp, {\it i.e.}, a kinematical effect, 
as it is located just at the threshold of $D$ plus $D^*$ mesons~\cite{Hanhart:2007yq,Aaij:2020qga}.
Yet it has been also claimed that $X(3872)$ is a superposition of a compact $c \bar c$ state and
a $D$-$\bar D^*$ molecular component~\cite{Takizawa:2012hy,Yamaguchi:2019vea}.
{This example shows that, in many cases, the interpretations are not conclusive 
due to uncertainties in hadron interactions and to the presence of many open channels.

An alternative picture for exotic resonances is a compact multi-quark (tetra-, or penta-quark) state.
QCD does not forbid such color-singlet multi-quark configurations. 
Indeed, many theoretical works predicted compact tetraquarks~\cite{Maiani:2004vq,Terasaki:2007uv,Takeuchi:2014uga}, 
pentaquarks~\cite{Jaffe:2003sg} and dibaryons~\cite{Jaffe:1976yi,Oka:1980ax}.
Yet, none of them has so far been confirmed experimentally, because the predicted states, that are
above some two-hadron thresholds, become resonances with often a large
fall-apart decay width.

Recently, with experimental developments in heavy hadron spectroscopy, 
possibilities of stable multi-quark states are being discussed frequently.
Let us focus on the simplest one, tetraquarks formed by two quarks and two antiquarks.
Compact tetraquarks may be composed of correlated colored diquarks
generated by the strong color Coulomb attraction.
It was suggested that this effect becomes critically important for systems with
two heavy quarks, $QQ^\prime \bar q \bar q^\prime$ in Refs.\cite{Karliner:2017qjm,Eichten:2017ffp}, 
where $Q^{(\prime)}$ and $q^{(\prime)}$ denote heavy ($c$ and $b$) and light ($u$, $d$, $s$) quarks, 
respectively.
Unlike the $Q \bar Q^\prime q\bar q^\prime$ system, 
$QQ^\prime \bar q \bar q^\prime$ is more likely to have
a bound state that is stable against strong decays,
mainly because the threshold energy for the latter, $Q\bar q + Q^\prime \bar q^\prime$, is
larger than the former, $Q\bar Q^\prime + q\bar q^\prime$.
In fact, there have been many theoretical studies about this possibility over the years
{(see, for instance, Ref.~\cite{Vijande:2009kj,Caramees:2018oue})}, which however remained inconclusive.
Meanwhile, the existence of the doubly charmed baryon $\Xi_{cc}$ has been experimentally 
established~\cite{Aaij:2018gfl}. 
This made a semi-quantitative discussion for double heavy teraquarks 
possible, giving large binding energies 
from an empirical mass formula~\cite{Karliner:2017qjm,Karliner:2014gca}.

The purpose of this paper is to systematically study stable $QQ^\prime\bar q\bar q^\prime$ 
tetraquark states 
with various flavor combinations in the non-relativistic quark model.  
We find several stable states, one of which is a strongly bound $bb \bar q \bar q$ 
with isospin and spin-parity $I(J^P) = 0(1^+)$, 
having a binding energy of almost  200 MeV.  
This confirms the earlier 
{discussions}~\cite{Karliner:2017qjm,Karliner:2014gca} and  is also consistent with the predictions of 
lattice QCD~\cite{Francis:2016hui,Junnarkar:2018twb,Hudspith:2020tdf,Mohanta:2020eed}.  
We have also found a shallow state for the same $I(J^P) = 0(1^+)$ channel.
By computing density distributions, it is shown that 
the deep one is a compact tetraquark state, while the shallow one is regarded as a 
loosely-bound molecule of two color singlet mesons, $B$ and $B^*$.
This is a hadronic analogue of the cluster formation in light nuclei~\cite{Ikeda:1968kk}, 
the first example that hadrons with totally different nature emerge from a single Hamiltonian.  
It is a universal feature of quantum many-body systems 
which will clarify unsolved problems of colored QCD dynamics.


For the quark model Hamiltonian, we employ the form of 
AP1 of Ref.~\cite{SilvestreBrac:1996bg} (See Eq.~(2) of \cite{SilvestreBrac:1996bg}), which is 
composed of a power-law confinement term and a gluon-exchange potential with non-relativistic kinetic
energy.
This Hamiltonian has been also employed for our former studies of pentaquarks of
$qqqc\bar c$ and $sss c \bar c$~\cite{Hiyama:2018ukv,Meng:2019fan}.
For determining the existence of bound states, it
is important for the calculation 
to treat the relevant threshold energies 
consistently. 
In order to improve the fit to the threshold meson masses, we
have tuned the potential parameters. 
In Table~\ref{table_AP1mesons}, we 
compile the values of the Hamiltonian parameters and the calculated masses
of the heavy mesons relevant 
to the present study of  tetraquarks. 
Compared with 
the experimental values, 
the meson masses are reproduced 
within the errors of at most 30 MeV or much less.  
The errors of the binding energies are expected to be less, 
as large part of errors will be cancelled
by taking the mass differences of the tetraquark and 
threshold mesons.

One missing element here is hadron dynamics, in particular meson-exchange 
interactions at long distances.
There are reasons, however, important features of our present discussions are robust.  
For deeply bound compact states such dynamics can be negligible.  
Whether or not shallow states exist may be modified, while their molecular 
structure remains unchanged as long as binding energies are small.

\begin{center}
\begin{table}[htb]
\caption{
The parameters of the Hamiltonian and the calculated masses (Cal) of heavy mesons compared with 
their experimental values (Exp).}
\label{table_AP1mesons}
\begin{tabular}{lrlr|rlrr}
\hline
\multicolumn{3}{c}{Parameters} &&& \multicolumn{3}{c}{Masses (MeV)}\\
&&&&&&Cal&Exp\\
\hline
$m_{u,d}$ (GeV) &&0.277&&& $\eta_b (0^-)$ &  \   9375 &\  9399 \\
$m_s$ (GeV)  &&0.593&&& $\Upsilon (1^-)$    &  9433 & 9460 \\
$m_c$ (GeV) &&1.826&&& $\eta_c (0^-)$   &  2984  & 2984\\
$m_b$ (GeV) &&5.195&&& $J/\psi (1^-)$    &  3102  & 3097\\
$p$ &&$2/3$&&& $B^- (0^-)$   &  5281  & 5279\\
$\kappa$ &&0.4222&&& $B^{*-} (1^-)$  & 5336  & 5325\\
$\kappa'$ &&1.7925&&& $B_s (0^-)$   & 5348  & 5367\\
$\lambda$ (GeV$^{5/3}$) &&0.3798 &&& $B_s^* (1^-)$  & 5410  & 5415\\
$\Lambda$ (GeV)&&1.1313&&& $D^- (0^-)$  & 1870  & 1870 \\
$A$ (GeV$^{B-1}$) &&1.5296 &&& $D^{*-} (1^-)$  & 2018 & 2010 \\
$B$ &&0.3263&&& &  & \\
\hline
\end{tabular}
\end{table}
\end{center}

{To} solve the four-body problem accurately, 
we employ the Gaussian expansion method \cite{Hiyama:2003cu}. 
The variational wave function of a tetraquark,  $\Psi_{I,JM}$, with isospin $I$ and total spin $(J,M)$ is formed 
as follows:
\begin{eqnarray}
	\Psi_{I, JM}=&\sum_C \xi_1^{(C)} \sum_{\gamma} B_{\gamma}^{(C)}\eta^{(C)}_I 
  \bigg[\Big[\big[[\chi_{\frac{1}{2}}\chi_{\frac{1}{2}}]_s \chi_{\frac{1}{2}} \big]_\Sigma 
 \chi_{\frac{1}{2}} \Big]_K \nonumber\\
	&\times  \big[[\phi^{(C)}_{n\ell}({\bf r}_{C}) \psi^{(C)}_{NL}({\bf R}_C)]_\Lambda
\phi'^{(C)}_{\nu \lambda }(\boldmath{\rho}_C) \big]_G \bigg]_{JM} ,
\label{total wave function}
\end{eqnarray}
where $\xi_1$ stands for the color singlet (indicated by the lower index 1) wave function, $\eta$ for the isospin of
light quarks, $\chi$ for the spin of each quark,
and $\phi$, $\psi$, $\phi'$ denote spatial wave functions.
The label $(C)$ specifies a set of Jacobi coordinates shown in Fig.~\ref{fig1_jacobi}, 
which are to coincide with the color combinations of quarks.
When two quarks are connected by a line, they form 
{a} color $\bar 3$, while a quark and an antiquark
will be connected to form a color singlet state. 
For example, the color wave functions, $\xi_1^{(C)}$, for $C=1$ and 4 are given by
$\xi_1^{C=1}=[((12)_{\bar 3}, 3)_3, 4]_1$ and $\xi_1^{C=4}=[((14)_{1}, 3)_{\bar 3}, 2]_1$, respectively.
The label $\gamma$ in Eq.~(\ref{total wave function}) includes all quantum numbers needed for the expansion, 
$\gamma \equiv \{s,\Sigma,K,n,N,\nu,\ell,L,\lambda, G\}$.

The expansion coefficients, or the variational parameters, $B_{\gamma}^{(C)}$, are determined 
by matrix diagonalization. 
Details of the method and its validity and accuracy are discussed in Ref~\cite{Hiyama:2003cu}. 
It should be noted that the precision is very important in the present analysis because the bound states are
often 
{close to} the two-body thresholds, where the system becomes very dilute, 
{making it} much
harder to obtain accurate wave 
{functions and eigenenergies}.

\begin{figure}
\centering
\includegraphics[width=0.5\textwidth]{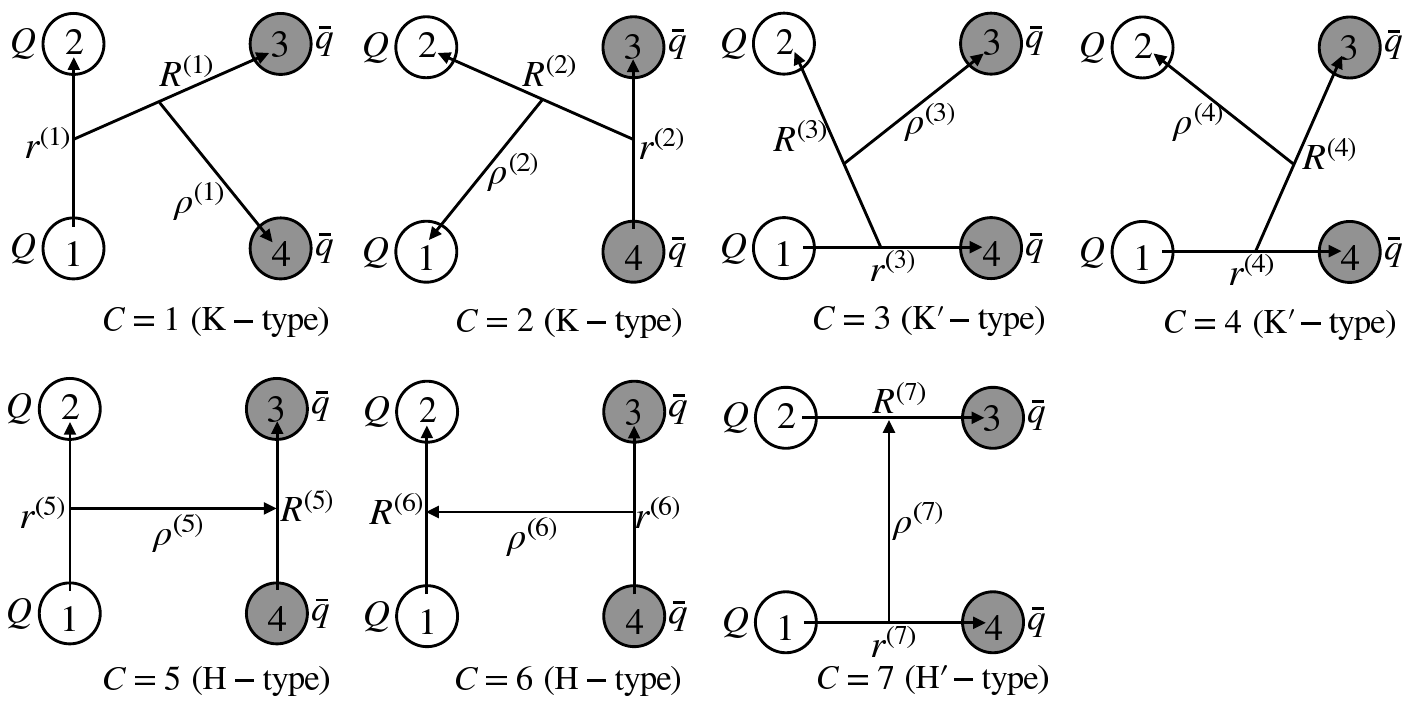}
\caption{ Seven sets of Jacobi coordinates for $QQ\bar q \bar q$ tetraquarks. 
The heavy quarks $Q$ and light anti-quarks $\bar q$ are labeled by 1, 2 and 3,4, respectively.  
They are classified into four types according to the color combinations, as $K$ ($C=1,2$), $K'$ ($C=3,4$), $H$ ($C=5,6$),  and $H'$ ($C=7$) types.} 
\label{fig1_jacobi}
\end{figure}

 
Bound tetraquark states, $Q Q^\prime \bar q\bar q^\prime$, are searched
for various flavor combinations from light to heavy quarks 
with spin and parity $J^P = 0^+, 1^+$ and $2^+$.
In the presence of light quarks, flavor combinations are expressed 
by isospin $I$.  
We have found altogether ten bound tetraquarks
as shown in Fig.~\ref{fig2_spectrum}, 
six for $J^P = 1^+$ (red bars), 
two for $0^+$ and two for $2^+$ (blue bars).
Other combinations, such as the one with all heavy quarks, do not 
accommodate stable states due to the relatively low threshold masses of fall apart mesons. 
We therefore conclude that the combination of heavy and light quarks is 
the key to generate stable bound states.

In Fig.~\ref{fig2_spectrum}, the resulting energies $-E_B$ 
($E_B$: binding energy) are shown in units of MeV 
together with their quantum numbers $I(J^P)$. 
In the figure, dashed bars stand for fall-apart two meson thresholds as indicated beside 
the bars. 
The columns are drawn relative to the threshold 
energies of 
the pseudoscalar ($0^-$) plus vector ($1^-$) 
meson masses such as 
$BB^*, DB^*$ for each quantum number.

\begin{figure}[htb]
\centering
\includegraphics[width=0.5\textwidth]{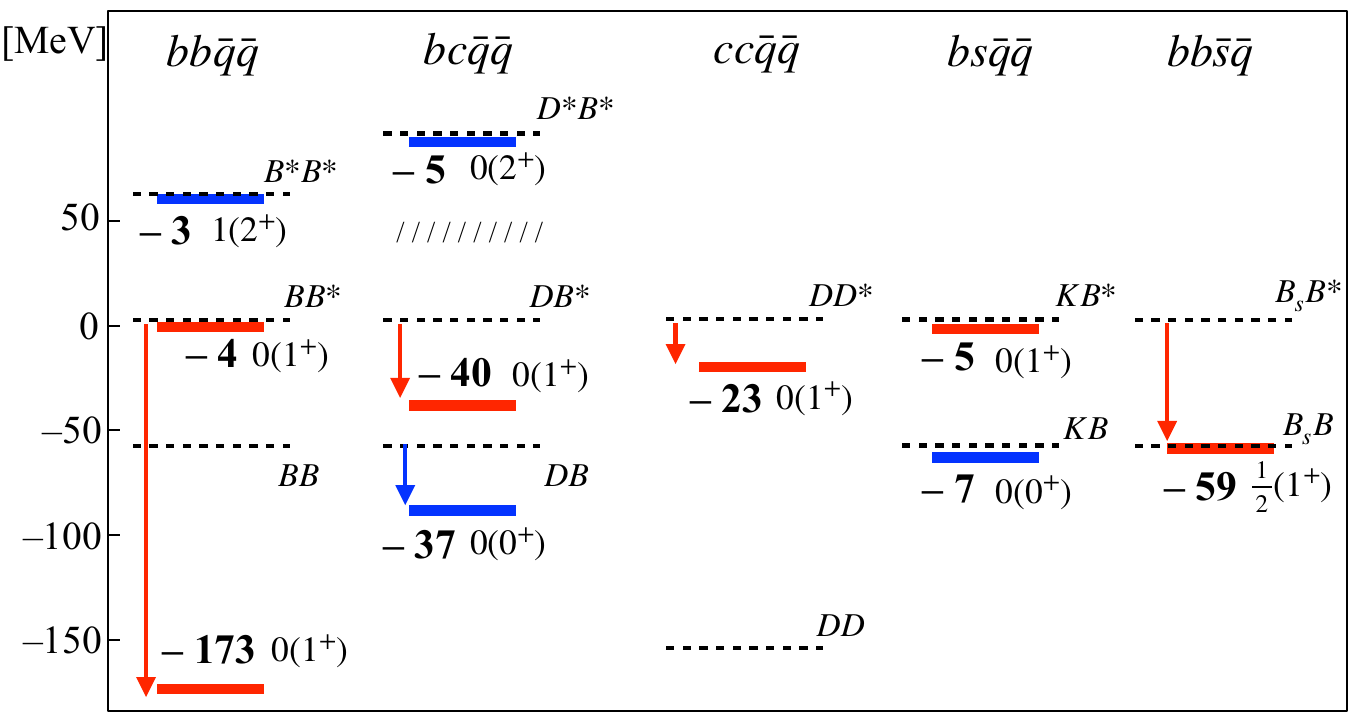}
\caption{Bound tetraquarks with their energies $-E_B$ (MeV) 
measured from the thresholds for various flavor contents. 
The labels 
beside each bar indicate isospin and spin-parity quantum numbers
$I(J^P)$.   The hatch pattern in the $bc\bar q \bar q$ sector indicates 
that the distance between the $DB^*$-$D^*B^*$ thresholds does not reflect 
the actual scale.
}
\label{fig2_spectrum}
\end{figure}

Let us discuss the nature of these bound states.  
\underline{$J^P = 1^+$}: 
For {$bb \bar u \bar d$ ($I=0$)}, we have obtained two bound states; 
one is deeply bound with a binding energy of 173 MeV, and 
the other shallow one with a binding energy of 4 MeV.  
As we will discuss shortly, these two states have very different internal structures.  
If we change the bottom quarks to charm or strange quarks 
for the deeply bound state, 
its binding energy decreases; 
specifically, in the order of the reduced masses of the quark pairs 
$bb$, 
$bc$, 
$cc$, 
$bs$, 
it decreases systematically as
173, 40, 23 and 5 MeV, respectively.  

This behavior is explained by the color electric force between heavy quarks,
as emphasized in Refs.~\cite{Karliner:2017qjm,Eichten:2017ffp}.  
For color $\bar 3$ states, it provides half of the attraction strength
of the color singlet quark and antiquark pair.  
Moreover, due to its $1/r$ behavior at short distances the attraction
increases proportional to the reduced mass of the two quarks.  
To {demonstrate} this explicitly, we plot the expectation values of 
the Coulomb ($1/r$) term of the color-electric potential for 
the $bQ$ pair in a $bQ \bar q \bar q$ tetraquark (red line) and for 
the $QQ$ pair in a $QQ \bar q \bar q$ tetraquark (blue line) 
as functions of $m_Q$ in Fig.~\ref{fig3_coulomb}.  
When $m_Q = m_b$, the two results agree, 
with the large attraction energy of $\sim  -200$ MeV.  
As $m_Q$ decreases down to $\sim 1$ GeV, 
where a bound state still exists, the absolute values of both the $QQ$ and $bQ$ energies 
decrease monotonically. 
The Coulomb energy for $bQ$ is more attractive than for $QQ$, 
because the reduced mass of $bQ$ is larger than that of $QQ$.  
The increase in the attractive energy is also understood intuitively by the decrease 
in the size of the $bQ$ pair as  shown in Table~\ref{table_msr}.  

\begin{figure}[htb]
\centering
\includegraphics[width=0.35\textwidth]{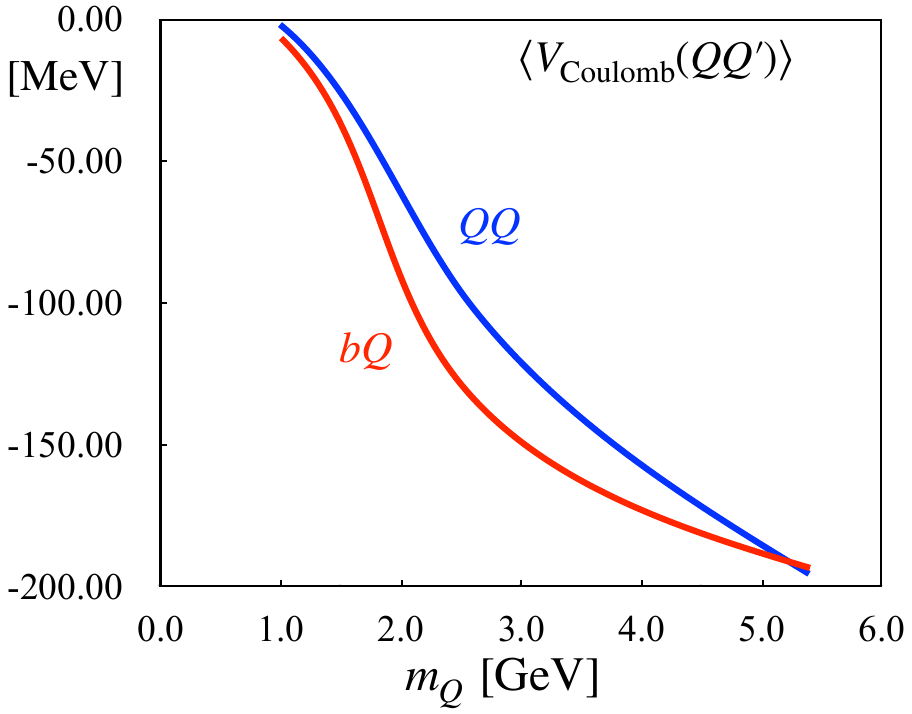}
\caption{Coulomb energies of the $bQ$ pair in the $bQ \bar q \bar q$ tetraquark (red line)
and that of the $QQ$ pair in the $QQ \bar q \bar q$ tetraquark (blue line), 
as functions of $m_Q$. }
\label{fig3_coulomb}
\end{figure}

There is another bound state for $bb \bar s \bar q: I(J^P) = 1/2(1^+)$ with 
a binding energy of 59 MeV.
This is the strange analogue of the deeply bound state of 173 MeV.
The difference between the two energies is 
partly due to the the spin-spin interaction, 
which is weaker for the strange quark than for the up and down quarks.  

\underline{\it Other $J^P$'s}: 
We have found two bound states with $I(J^P) = 0(0^+)$ for $bc \bar q \bar q$ 
bound below the $BD$ threshold by 37 MeV, and for $bs \bar q \bar q$ by $7$ MeV.  
Their $Q$ and $Q'$ are in symmetric configurations, 
so that their siblings in the $bb\bar q \bar q$ or $cc\bar q \bar q$ channels are forbidden by the Pauli principle.
This is realized in a lattice QCD calculation as well~\cite{Junnarkar:2018twb}.

Lastly, we have also found two more states with $J^P = 2^+$.
The one in the $bb \bar q \bar q$ channel of $I=1$ is located only 3 MeV below the $B^*B^*$ threshold.
This state is formed by the bad  anti-diquark $\bar q \bar q$ of ($I(J^+)=1(1^+)$) bound to 
the heavy vector diquark
$bb$.
The mass difference from the state of $0(1^+)$ with 173 MeV binding energy 
can mostly be explained by the spitting between the good ($I(J^P) = 0(0^+))$ and bad 
anti-diquarks.
The other $2^+$ bound state 
appears in a $bc \bar q \bar q$ configuration with a small binding energy 
of 5 MeV below the $D^*B^*$ threshold.


\begin{center}
\linespread{1.2}
\begin{table}[htb]
\caption{The energies of stable tetraquarks $-E_B$ in comparison with recent lattice QCD calculations 
in units of MeV~\cite{Francis:2016hui,Junnarkar:2018twb,Hudspith:2020tdf,Mohanta:2020eed}. 
N stands for ``no bound state''.
Refs.~\cite{Francis:2016hui,Junnarkar:2018twb,Mohanta:2020eed}  
report binding energies directly, 
while only the meson and tetraquark energies are given in Ref.~\cite{Hudspith:2020tdf}. 
The shown errors for Ref.~\cite{Hudspith:2020tdf} here are estimated by combining the errors of the individual hadron masses in quadrature.
}
\label{table_result_comparison}
\begin{tabular}{llccccc}
\hline
&  $I(J^P)$& {This work} & \cite{Francis:2016hui}   & \cite{Junnarkar:2018twb}  &  \cite{Hudspith:2020tdf} &  \cite{Mohanta:2020eed}  \\
\hline
$bb\bar{q}\bar{q} $ & $0(1^+)$ &$-173$&$-189\pm13$&$-143 \pm 34$ & $-$ & $-186 \pm 15$\\
$bc\bar{q}\bar{q} $ & $0(1^+)$ & $-40$ & $-$ & $-$ & $13 \pm 3$& $ -$ \\
$cc\bar{q}\bar{q} $ & $0(1^+)$& $-23$ & $-$ &$-23 \pm 11$& $-$ & $-$ \\
$bs\bar{q}\bar{q} $ & $0(1^+)$ & $-5$ & $-$ & $-$ & $16 \pm 2$ & $-$ \\
$bb\bar{s}\bar{q} $ & $\frac{1}{2}(1^+)$ & $-59$ & $-98 \pm 10$ & $-87 \pm 32$ & $-$ & $-$\\ 
\hline
$bb\bar{q}\bar{q} $ & $1(0^+)$ & N & $-$ & $-5 \pm 18$ & $-$ & $-$ \\
$bc\bar{q}\bar{q} $ & $0(0^+)$ & $-37$ & $-$ & $-$ & $17 \pm 3$ & $-$ \\
$cc\bar{q}\bar{q} $ & $1(0^+)$ &  N & $-$ & $26 \pm 11$ & $-$ & $-$ \\
$bs\bar{q}\bar{q} $ & $0(0^+)$ & $-7$ & $-$ & $-$ & $18 \pm 2$ & $-$ \\
\hline
\end{tabular}
\end{table}
\end{center}

Next, we compare our results with those of recent lattice QCD 
calculations~\cite{Francis:2016hui,Junnarkar:2018twb,Hudspith:2020tdf,Mohanta:2020eed} 
in Table~\ref{table_result_comparison}. 
We see that for the channels containing either $cc$ or $bb$ heavy quarks, the agreement between the lattice and our 
quark model results is rather good. 
Especially for the deeply bound $bb\bar{q}\bar{q}$ and $bb\bar{s}\bar{q}$ cases with $I(J^P) = 0(1^+)$ for which 
calculations of multiple lattice QCD collaborations are available, 
the quark model states lie within an energy range of at most 40 MeV of the lattice results. 
For all other states with $cc$ or $bb$ heavy quarks, the bound states, if any, are only rather shallow both for the quark model and the lattice calculations. 
Conversely, for the channels with $bc$ and $bs$ quarks which have 
been studied in Ref.~\cite{Hudspith:2020tdf}, there is some disagreement 
between the lattice and the quark model results. 
Specifically, we find bound states for all of them in our work, 
while on the lattice no such bound state is obtained. 


We continue by discussing the two-body density distributions for 
quark pairs in the tetraquarks, 
which will help 
revealing their
spatial structure. 
The two-body density distribution
of a $qq^\prime$ pair,
where $q$ or $q^\prime$ indicates any quark 
or anti-quark in the tetraquark,
 is defined by 
\begin{equation}
\rho_{qq^\prime}(r_{qq^\prime}) = \int d\hat{\bm r}_{qq^\prime} d{\bm x}_1 d{\bm x}_2 \, |\Psi_{JM}({\bm r}_{qq^\prime}, {\bm x}_1,{\bm x}_2)|^2
\end{equation}
where $r_{qq^\prime}=|{\bm r}_{qq^\prime}|$ is the distance between 
$q$ and $q^\prime$, $\hat{\bm r}_{qq^\prime}$ is
the angular part of the relative $q$-$q^\prime$ coordinate, 
and ${\bm x}_1$ and ${\bm x}_2$ denote the other Jacobi coordinates.

In Fig.~\ref{fig4_density}, we show $r^2 \rho_{qq^\prime}(r)$ 
for various $qq^\prime$ pairs in the two $bb \bar q \bar q$ tetraquarks of $I(J^P) = 0(1^+)$.  
For the deeply bound state (a), we see a very compact structure for the $bb$ pair, 
while the $b \bar q$ and $\bar q \bar q$ pairs have extended density distributions.
This is what we expect; the $bb$ pair 
is strongly attracted due to the color-electric force,
while this effect is smaller for the $bq$ and $qq$ pairs as the attraction is proportional to their reduced masses.  
Turning to the 
shallow bound state (b), all diquark pairs are 
extended and furthermore, 
the $bb$ distribution shows a node-like structure. 
This implies that this state is a nodal excitation of the $bb$ pair.

To understand these features 
more quantitatively, we summarize in Table~\ref{table_msr}, the mean distances,
$R_{qq^\prime}\equiv  \left( \int r^2 \rho_{qq^\prime}(r)\,r^2 dr/ \int \rho_{qq^\prime}(r)\,r^2 dr\right)^{1/2}$, 
of various pairs of quarks (and antiquarks).  
One sees clear tendencies that the density distributions depend on the types of quark pairs and their binding energies. 
Namely, the deep bound states have a smaller 
$R_{Q\bar q - Q^\prime \bar q}$, the distance between the centers of mass of 
$Q\bar q$ and $Q^\prime \bar q$, compared to the shallow ones, for which
$R_{Q \bar q}<R_{Q\bar q - Q^\prime \bar q}$.
This indicates that the shallow states are loosely bound (molecular) states of color singlet 
mesons, $(Q \bar q)_1 + (Q^\prime \bar q)_1$, 
where the index 1 denote color singlet.  
In particular, the node-like structure of $bb$ may transfer to the similar structure 
for the mesons.  
It is very interesting to see two extreme cases of bound states,
one deep and compact, the other shallow and molecular, simultaneously
in the spectrum of the single quark model Hamiltonian.
This is the first example of a hadronic analogue of cluster formation in spectra of light nuclei, 
where cluster structures made of $\alpha$ particles are developed 
around the $\alpha$ emission thresholds~\cite{Ikeda:1968kk},
while the lower bound states are compact shell-model-like states.

\begin{figure}[htb]
\centering
\includegraphics[width=0.5\textwidth]{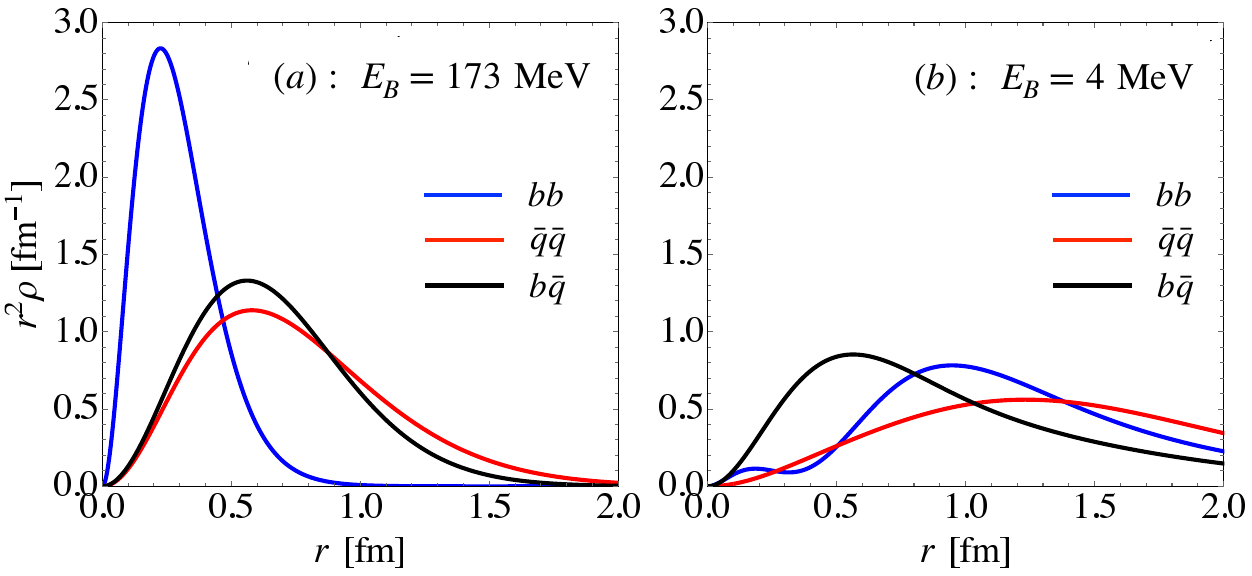}
\caption{Density distibutions for various quark pairs in the deep (a) and shallow (b) $bb  \bar q \bar q$ tetraquarks of $J^P = 1^+$.  }
\label{fig4_density}
\end{figure}

\begin{center}
\linespread{1.2}
\begin{table}[htb]
\begin{tabular}{c c c c c c c c }
\hline
 $QQ^\prime \bar q \bar q$   & $I(J^P)$ & \  $-E_B$   \  &  
 \ $R_{QQ^\prime}$ \ &  \ $R_{Q \bar q}$ \ & \ $R_{Q^\prime \bar q}$ \ & \ $R_{\bar q \bar q}$  \ &
$R_{Q\bar q - Q^\prime \bar q}$\\
\hline
$bb \bar q \bar q$  &  $0(1^+)$ & $-173$ & 0.34 & 0.84         &        & 0.74     & 0.32 \\
$bb \bar q \bar q$  &  $0(1^+)$ & $-4$ & 1.09  & 0.93          &               &  1.11    &  1.07\\
$bc \bar q \bar q$  &  $0(1^+)$ &$-40$ & 0.65  & 0.79          & 0.80                & 0.94 & 0.61 \\
$cc \bar q \bar q$  &  $0(1^+)$ & $-23$ & 0.83  & 0.85          &                &  1.00    & 0.75 \\
$bc \bar q \bar q$  &  $0(2^+)$ & $-5$ & 1.72  & 1.38          &    1.40    &  1.93   & 1.57 \\
\hline
\end{tabular}
\caption{Mean distance $R_{qq^\prime}$ [fm] for various tetraquarks.
Binding energies $E_B$ are in units of MeV. }
\label{table_msr}
\end{table}
\end{center}

The states that we have discussed so far are stable against the strong decay, while
they  decay through the electro-magnetic or weak interactions. 
For example, the $I(J^P) = 0(1^+)$ state of $bc \bar u \bar d$ with binding energy 40 MeV 
will decay radiatively into $D+B+\gamma(M1)$.  
Similarly all 
the $J^P = 1^+$ states above the two $0^-$ meson 
thresholds, 
and $2^+$ states above the $0^-$ and $1^-$ meson 
thresholds, 
are subject to such decays.  
The two 
deeply bound states, 
{the} $bb \bar u \bar d$ ($0(1^+)$) 
and $bc \bar u \bar d$ ($0 (0^+)$)
states, 
on the other hand, can decay only via the weak interaction.

Summarizing, we have found a few stable bound states in $QQ'\bar q\bar q'$ tetra quark systems
in the quark model. 
The deep compact bound state in $bb\bar u\bar d$ (and also in $cc\bar u\bar d$) with $I(J^P)=0 (1^+)$ 
agrees well
with the lattice QCD prediction. 
A shallow $bb\bar u\bar d$ ($0(1^+)$) bound state is also found, whose wave function is consistent with a molecule-type loosely bound state of $B$ and $B^*$ mesons.
This is the first hadronic example of a set of a deep and shallow bound states in the same channel.

\noindent
{\bf Acknowledgments:}\\ 
This works is supported in part by  Grants-in Aid for Scientific Research on Innovative Areas, No. 18H05407 for QM, EH, AH, and JP19H05159 for MO.  
KUC is supported by the Australian Research Council Grant DP190100297. 
P.G. is supported by the Grant-in-Aid for Early-Carrier Scientists (JP18K13542), 
Grant-in-Aid for Scientific Research (C) (JP20K03940) 
and the Leading Initiative for Excellent Young Researchers
(LEADER) of the Japan Society for the Promotion of Science (JSPS).

\bibliography{apssamp}

\end{document}